\documentstyle[12pt,epsf]{article}
\newcommand{\barr}{\begin{array}}
\newcommand{\earr}{\end{array}}
\newcommand{\beq}{\begin{equation}}
\newcommand{\eeq}{\end{equation}}
\newcommand{\bea}{\begin{eqnarray}}
\newcommand{\eea}{\end{eqnarray}}

\newcommand{\prlr}[3]{Phys.\ Rev.\ Lett.\ {\bf #1}, #2 (#3)}

\newcommand{\prbr}[3]{Phys.\ Rev.\ B~{\bf #1}, #2 (#3)}
\newcommand{\prdr}[3]{Phys.\ Rev.\ D~{\bf #1}, #2 (#3)}
\newcommand{\plbr}[3]{Phys.\ Lett.\ B~{\bf #1}, #2 (#3)}
\newcommand{\plar}[3]{Phys.\ Lett.\ A~{\bf #1}, #2 (#3)}
\newcommand{\npbr}[3]{Nucl.\ Phys.\ {\bf #1}, #2 (#3)}

\newcommand{\jpcr}[3]{J.\ Phys.\ {\bf #1}, #2 (#3)}

\makeatletter

\def\compoundrel#1\over#2{\mathpalette\compoundreL{{#1}\over{#2}}}
\def\compoundreL#1#2{\compoundREL#1#2}
\def\compoundREL#1#2\over#3{\mathrel
	{\vcenter{\hbox{$\m@th\buildrel{#1#2}\over{#1#3}$}}}}
\makeatother

\def\c+{c^{\dagger}}
\def\d+{d^{\dagger}}

\include{psfig}

\begin{document}

\title{Infra-Red Stable Supersymmetry in Chern-Simons Theories with Matter and 
Quenched Disorder}
\author{H.~Hamidian}
\date{\today}

\maketitle
\begin{center}
{\sl Department of Physics, Stockholm University, Box 6730, S-113~85 Stockholm, 
Sweden}
\end{center}

\begin{abstract}

\thispagestyle{empty}

We study Abelian Chern-Simons field theories with matter fields and global 
$SU(N)$ symmetry in the presence of random weak quenched disorder. In the 
absence of disorder these theories possess ${\cal N}=2$ supersymmetric fixed 
points and ${\cal N}=1$ supersymmetric fixed lines in the infra-red limit. We 
show that although the presence of disorder forbids any
supersymmetry of the bare action, infra-red stable supersymmetric fixed points 
and fixed lines are realized in the disorder-averaged effective theories.
\\
\\
\\
\noindent PACS number(s): 11.15.-q, 11.10.Kk, 11.30.Pb, 11.10.Hi, 11.30.Qc
   
\end{abstract}

\newpage
Model systems in less than four (space-time) dimensions have often played an 
important role in understanding general methods and applications of quantum 
field theory and developing ideas in four dimensions.

Notable among these are Chern-Simons (CS) gauge field theories in (2+1) 
dimensions \cite{jackiw1}, which have a long history and have been 
intensively studied in recent years. There are 
several motivations for studying CS theories. From a purely theoretical and 
mathematical point of view they constitute a new class of (topological) gauge 
theories and are closely related to certain models in topological quantum field 
theory 
\cite{schwartz} in three space-time dimensions and integrable statistical 
mechanical models and rational conformal field theories in two dimensions 
\cite{moore}. From the more physical point of view there has been general 
interest in CS theories due to their resemblance to the high-temperature limit 
of (3+1) dimensional gauge theories and their possible relevance to certain 
condensed matter phenomena, in particular high-$T_C$ superconductivity 
\cite{wen1} and the quantum Hall effect \cite{block}.

Although CS gauge theories are in many respects similar to their 
four-dimensional counterparts, there are substantial differences which are 
intimately 
connected with the appearance of the CS three-form in the action \cite{siegel} 
and parity anomalies \cite{niemi}. Formally, CS gauge theories are strictly 
renormalizable quantum field theories and their perturbative expansions contain 
logarithmic divergences. However, contrary to the Yang-Mills and Maxwell gauge 
theories, and due to their topological nature, infinite charge renormalizations 
are absent in both Abelian and non-Abelian CS 
gauge theories \cite{coleman1,guadagnini}. 

The situation changes significantly when matter fields are added to the CS 
theory. Although the CS gauge coupling remains unrenormalized as before, the 
matter couplings require renormalization. This leads to non-vanishing 
$\beta$-functions exhibiting a complicated and non-trivial renormalization group 
(RG) flow and a rich set of fixed points \cite{grigoryev}. By analyzing the CS 
gauge theory with matter couplings in the infrared (IR) limit the authors of 
Ref. \cite{grigoryev} were able to show that, quite unexpectedly, some of 
the resulting fixed points and fixed lines exhibit ${\cal N}=1$ and ${\cal N}=2$ 
supersymmetry, 
which are realized as IR stable or saddle point solutions of the corresponding 
RG equations. 

In this letter we study Abelian CS field theories with matter in the presence of 
random weak quenched disorder (QD) at zero temperature. The model is described 
by the (Euclidean) action,
\bea
S[\delta r] &=& \int d^3x \Big[ 
\frac{i}{2} \epsilon^{\mu \nu \lambda} A_\mu \partial_\nu 
A_\lambda + |D_\mu \phi_j|^2 + i {\bar \psi_j} {\not \!\! D} \psi_j
\nonumber\\&&
+ (m^2 + \delta r(x)) \phi^*_j \phi_j - M {\bar \psi_j} \psi_j 
\nonumber\\&&
- q(\phi^*_j \phi_j)^2
- \alpha {\bar \psi_j} \psi_j \phi^*_k \phi_k - \beta {\bar \psi_j} \psi_k 
\phi^*_j \phi_k 
\nonumber\\&&
- \frac{1}{4} \eta ({\bar \psi_j} \psi^*_k \phi_j \phi_k + 
{\bar \psi^*_j} \psi_k \phi^*_j \phi^*_k) + h(\phi^*_j 
\phi_j)^3 
\Big],
\label{lagrangian1}
\eea 
where $D_\mu=\partial_\mu - i e A_\mu$ is the covariant derivative  
and summation over repeated indices is implicitly assumed.
Our notation and conventions are those of \cite{grigoryev,gates}.
The matter fields are in the fundamental representation of the global $SU(N)$ 
symmetry group $(j,k = 1,\ldots,N)$ and   
the spinors are defined in the Majorana basis and under charge 
conjugation $C$ transform according to $C:\psi \rightarrow \psi^*$ [Note that 
Yukawa type vertices, which are in general allowed in three dimensions, are 
excluded from (\ref{lagrangian1}) since they violate fermion-number 
conservation]. The QD is described by random 
fluctuations of the disorder (field) $\delta r(x)$ whose probability 
distribution 
is taken to be symmetric and Gaussian:
\bea
P[\delta r(x)] = p_0 \exp \left \{ -\frac{1}{4 \Delta} \int d^3x[\delta r(x)]^2 
\right \},
\label{gaussian}
\eea
with average fluctuations of width $\Delta \ll 1$,
\bea
\langle \delta r(x) \delta r(x') \rangle_{P[\delta r]} = \Delta \delta^3(x-x'),  
\label{gaussian2}
\eea
and $p_0$ the normalization constant.

In the absence of QD, 
i.e. when $\delta r(x) = 0$ in (\ref{lagrangian1}), and for the particular 
choice of the couplings,
\bea
\alpha &=& \lambda~,~\beta = \lambda + e^2~,~\eta = 2(\lambda - e^2),
\nonumber\\
h &=& \lambda^2~,~m = M~,~q = 2 \lambda m,
\label{couplings}
\eea
the theory possesses ${\cal N}=1$ and, upon imposing the further restriction 
$\lambda = e^2$, ${\cal N}=2$ supersymmetry; both of which are preserved by 
renormalization. 
By examining the RG equations for this theory the authours of 
Ref.~\cite{grigoryev} were able to establish the existence of ${\cal N}=1$ and 
${\cal N}=2$ supersymmetric IR fixed points. The ${\cal N}=2$ supersymmetric 
fixed ponit exists for any choice of the global $SU(N)$ group. For $N=1$ the 
fixed point is IR stable and the critical exponents are Gaussian, while for 
$N>1$ the fixed points are saddle point solutions of the RG equations. The 
${\cal N} = 1$ supersymmetry manifests itself as a fixed line and is also valid 
for any value of $N$. The only fixed point on this line is the ${\cal N} = 2$ 
supersymmetric one and is IR stable along the line, i.e. when the ${\cal N} = 1$ 
supersymmetry is imposed.  

The oiginal motivation that led us to consider the model described by (\ref{lagrangian1}) 
was to examine whether it is possible to modify the Abelian CS theory in such a way that 
the bare 
action 
would no longer support supersymmetry but, in the IR limit, the effective theory would flow 
towards 
stable supersymmetric fixed points and fixed lines as in the CS theory without QD. As will 
be 
discussed below, the inclusion of weak QD does indeed lead to such a possibility and could, 
in fact, 
arise in certain physically interesting theories. Among other applications, as mentioned 
above, the CS 
field theories have been considered as effective long-range theories of anyon 
superconductivity. In 
this 
context the CS fields arise from the Hopf term of an associated non-linear $\sigma$ model 
which 
describes 
spin excitations in doped quantum  antiferromagnets \cite{wen2} and it has been argued 
\cite{wen1,wen2} 
that a phase transition could occur which would lead either to superfluidity or, when the 
Maxwell field 
is introduced, to superconductivity. In studying such phase transitions the sytems are 
usually assumed 
to 
be perfectly homogeneous. However, in real samples defects and impurities are always 
present and can 
lead to different (critical) behavior. One can analyze the 
effect of impurities by using a field-theoretical description of phase transitions based 
on an effective Landau-Ginzburg formulation by introducing weak QD as, e.g., has been done 
here for the 
Abelian CS theory with matter. In other contexts, it may be possible to speculate on the 
role played 
by 
(topological) defects such as monopoles and strings in higher-dimensional theories, as long 
as these 
defects are 
sufficiently heavy and diluted and can be considered as effectively forzen-in on the scale 
of the entire system. In this case also one could imagine the system as described by an 
effective 
Landau-Ginzburg theory with weak QD which resembles the model studied here.         

We shall now proceed to study the model described by (\ref{lagrangian1}) and explore 
whether there 
exist supersymmetric IR fixed points in the presence of weak QD and, if there 
are such fixed points, whether they are stable against QD. 

In order to understand the effect of disorder one often invokes the so-called 
Harris citerion \cite{harris}. 
According to this criterion, a change in the critical behavior due to disorder 
can occur only if the specific heat exponent $\alpha$ of the theory without 
disorder (the {\em pure} system)  
is 
positive. The physical reasoning which leads to this criterion is as follows: In 
the familiar case of theories with only a single mass parameter (e.g. the 
Landau-Ginzburg theory describing the 3D Ising model) criticality 
sets in at $T=T_C$, where one identifies the temperature with the mass term in 
the Lagrangian as $m^2_{\rm bare} \sim t = (T-T_C)/T_C$. At 
the critical point the (renormalized) mass term vanishes and the correlation 
length $\xi \sim m^{-1}$ diverges like $(T-T_C)^{-\nu}$, for some exponent 
$\nu$. Writing $\Delta T \equiv T - T_C \sim \xi^{-1/\nu}$ and noting that the 
(statistical) fluctuations $\delta T_C$ in $T_C$ due to QD obey $\delta T_C \sim 
\xi^{-D/2}$, Harris has argued \cite{harris} that a sharp (second-order) 
transition is consistent if and only if the fluctuations in $T_C$ approach zero 
faster than does $\Delta T$ as $\xi$ approaches infinity. This occurs only when 
$\delta T_C/\Delta T = \xi^{(1/\nu - D/2)} \rightarrow 0$ as $\xi \rightarrow 
\infty$, or $2 - \nu D <0$. In other words, when $\alpha = 2 - \nu D < 0$ the QD 
does not affect the critical behavior and one expects only second-order phase 
transitions (although 
treatments using the $\epsilon$-expansion suggest that IR stability is, in 
general, endangered by long-range quantum fluctuations \cite{hamidian}), whereas 
$\alpha > 0$ indicates that including QD may either lead to (possibly new) IR 
stable fixed points or drive a smeared (first-order) phase transition.

As a first step to examine the effect of disorder on the critical behavior in the manner 
described 
above 
let us consider the critical exponent $\nu$ in the absence of QD which is given by the 
well-known 
relation \cite{zinn-justin} $\nu^{-1} = 2 - \gamma_{\phi^2}(g^*)$, where 
$\gamma_{\phi^2}(g^*)$ is the 
anomalous 
dimension of the $\phi^2$ operator at the fixed point $g^*$. The anomalous dimension at the 
critical 
point $g^*$ can be computed by using the RG equations for the Abelian CS theory without QD 
\cite{grigoryev} and vanishes as $g \rightarrow g^*$ in the IR limit. [Note that it is 
always possible 
to renormalize at zero momentum in any dimension $d \leq 4$ and define renormalized 
$n$-point 
correlation 
functions, $\Gamma^{(n)}_{\rm r}(p;m_{\rm r},g_{\rm r}) = m_{\rm r}^2 + p^2$, whose 
zero-momentum limit 
is relevant to the study of scaling behavior at criticality.] One therefore obtains $\alpha 
= 1/2$ for 
the Abelian CS field theory with matter (and without QD) which, 
depending on the value of   
${\cal N}$, corresponds 
to a Gaussian fixed point or fixed line. Since $\alpha > 0$ one may expect the 
QD 
to somehow affect the critical behavior. However since the supersymmetric IR 
stable 
theory exhibits Gaussian critical behavior with $q=m=M=0$ at 
criticality, a closer examination is required. This is due to the existence of 
the additional mass parameters, $q$ and $M$, in the theory which lead to 
further conditions that have to be satisfied in order for the critical theory to 
be 
well-defined in the IR limit. In fact, although the dimensionless couplings in 
the theory simply approach constant values at the IR fixed points, the RG 
equations demand that both $q$ and $M$ vanish at the critical point. To see this 
it is sufficient to simply note that, as a result of the RG equations, the 
dimensionless quantities ${\bar q} = q/\mu$ and ${\bar M} = 
M/\mu$ ($\mu$ is the RG scale) behave as
\bea
{\bar q}^2(p) \sim q_0^2/p^2,~~~{\bar M}^2(p) \sim M_0^2/p^2,
\label{RG1}
\eea
to the leading order in the $p^2 \rightarrow 0$ limit, which diverge unless also 
$q$,$M \rightarrow 0$ in 
the same limit. Therefore the low-energy (tri-)critical theory is consistently 
defined 
as long as not only $m$, but also $q$ and $M$ approach zero in the IR limit. By 
examining the $\beta$-functions in the theory without QD it can be easily 
confirmed that this is indeed the case \cite{grigoryev}. By adding QD to the 
theory, as described by (\ref{lagrangian1})-(\ref{gaussian2}), a new 
dimensionful 
parameter, $\Delta$, is introduced which, on dimensional grounds, can be  
expected to affect the equations describing the RG flows of all the dimensionful 
couplings in the theory. Similarly to ${\bar q}$ and ${\bar M}$, by defining the 
dimensionless quantity 
${\bar \Delta} = \Delta/\mu$ it follows that with the presence of QD a 
consistent (multi-)critical theory in the $p^2 \rightarrow 0$ limit exists only 
if the QD parameter $\Delta$ (effectively) renormalizes to zero at the IR fixed 
point. To check whether this is realized and, if so, whether the IR fixed points 
remain stable against (quantum) fluctuations due to QD requires the knowledge of 
the RG equations that govern the flow towards the fixed points and fixed lines.   
     
To derive the RG equations for the theory defined by (\ref{lagrangian1}) we must 
first compute the effective Landau-Ginzburg action by taking the average over 
the disorder $\delta r(x)$. This is 
most conveniently accomplished by using the standard 
replica trick \cite{edwards} which enables one to compute the disorder-averaged 
effective action by using the identity $\ln Z = \lim_{n \rightarrow 0} (Z^n - 
1)/n$, where $Z$ is the partition function. Using the probability distribution 
(\ref{gaussian}), 
we obtain the (replica) action,
\bea
S_{QD} = &&\sum_{a=1}^n S_a[0] - \int d^3x \int d^3x' \delta^3(x-x') 
\nonumber\\
&&\sum_{a,b=1}^n q_{ab} [\phi_i^a (x)]^2 [\phi_j^b (x')]^2.
\label{sqd}
\eea
Here $a,b = 1,\ldots , n$ are the replica indices, $q_{ab} = q \delta_{ab} - 
\Delta$ and each replica $S_a[0]$ is of the same 
form as in (\ref{lagrangian1}) with the same replica index $a$ for all the 
fields. It is then straightforward to compute the RG equations by using the 
Feynman rules derived from (\ref{sqd}) and taking the limit $n \rightarrow 0$. 
Since the theory is defined in three dimensions there are no genuine ultraviolet 
divergences in the one-loop approximation and the first non-trivial 
contributions arise from the 
two-loop diagrams. To this order, the $\beta$-functions which include the 
parameter $\Delta$ are
\bea
\beta_{\Delta} &=& \Big\{ \frac{1}{3} \big[ 20 N \alpha^2 + 40 \alpha \beta + 4 
(2 N + 
3) \beta^2 + (2 N + 5) \eta^2 - 40 (N +2) e^4 \big] + 24 (N + 2) h \Big\} 
\Delta,
\nonumber\\
\beta_{m^2} &=& \frac{1}{3}  \Big[ 4 (N \alpha^2 + 2 \alpha \beta + N \beta^2) + 
(N + 1) \eta^2
- 4 (5 N + 4) e^4 \Big] m^2 
\nonumber\\&&
- \Big[ 4 (N \alpha^2 + 2 \alpha \beta + N 
\beta^2) + (N + 1) \eta^2 - 4 N e^4 \Big] M^2 
\nonumber\\
&& + 4 (N + 1) (q - \Delta)^2 + 4 N \Delta (q - \Delta),
\label{betas}
\eea     
where
$\beta_\Delta=\mu^2 d \Delta / d \mu^2$, $\beta_{m^2} = \mu^2 d m^2 / d \mu^2$, 
and 
we have omitted the familiar factors of $(64 \pi^2)^{-1}$. The $\beta$-functions 
for the dimensionless couplings, $e, \alpha, \beta, \eta$ and $h$ 
and the dimensionful parameters $q$ and $M$ are 
not affected by QD to the order considered here and their corresponding 
expressions, which are somewhat lengthy, can be found in \cite{grigoryev}. 

The fixed points correspond to the zeros of the $\beta$-functions for the 
dimensionless couplings. The CS gauge coupling $e$ has a vanishing 
$\beta$-function to all orders in perturbation theory and can be taken as a 
fixed parameter, while the other couplings become proportional to powers of $e$ 
with coefficients that are functions of $N$. Using the $\beta$-functions in 
(\ref{betas}) we find that the IR stable fixed points occur at $\Delta = 0$. 
In particular, this implies that although the theory defined by 
(\ref{lagrangian1}) cannot support supersymmetry, the weak QD vanishes as the  
the IR stable ${\cal N}=1$ and ${\cal N}=2$ supersymmetric fixed lines and fixed 
points are approached and the     
Gaussian critical behavior is recovered. This non-trivial result provides an 
example 
for the existence of gauge theories with matter in which the (bare) Lagrangian 
does not possess supersymmetry but the theory flows towards a stable 
supersymmetric fixed point in the IR limit. Although we shall not present it 
here, we have checked that the same results are also true for certain 
non-Abelian CS field theories with matter. Non-Abelian CS field theories with 
matter with $SU(n), Sp(n)$ and $SO(n)$ gauge groups were studied in 
\cite{grigoryev1} where similar behavior to the Abelian CS field theories were 
obtained. Adding QD to such theories leads to the same conclusions as for the 
Abelian theory discussed here. Although the Abelian and the non-Abelian 
theories exhibit quite different features---e.g. with respect to the dependence 
of the IR stability of the fixed points on the parameter $N$ of the global 
$SU(N)$ group---they both behave similarly with regard to the restoration of 
supersymmetry at the IR stable fixed points when QD is present.

To conclude, in this letter we have studied Abelian CS field theories with 
matter in the presence of random weak QD. We have 
shown that although in the presence of disorder these theories cannot support 
any
supersymmetry of the bare action (unlike such theories without QD), stable 
supersymmetric fixed points are realized in the infra-red 
limit of the disorder-averaged effective theories. Several issues need to be 
explored in connection with this non-trivial behavior in three dimensions. An 
important phenomenological question that arises in the study of supersymmetric 
theories is to understand the origin of supersymmetry breaking. There are three 
main approaches to answer this question, namely: the introduction of explicit 
soft supersymmetry breaking terms, supersymmetry breaking in a hidden sector 
(such as the one based on ${\cal N}=1$ supergravity with $F$ and $D$ components 
of some chiral and vector superfields obtaining vev's at typically high energy 
scales) and supersymmetry breaking at low energies through interactions with the 
gauge fields \cite{dimopoulos}. All these approaches, however, still suffer from 
important drawbacks (see Dine in Ref. \cite{dimopoulos} for a brief review and 
further references) and there is much room for exploring other possibilities. In 
the three-dimensional model presented in this letter we have shown that it is 
possible to start with certain non-supersymmetric Lagrangians at some 
high energy scale, while maintainig supersymmetry in the low-energy effective 
theory. The supersymmetry of this effective theory can itself be broken by some 
appropriate mechanism to obtain another effective theory which is not 
supersymmetric. It is thus possible to imagine an intermediate supersymmetric 
theory between two non-supersymmetric theories in this way. The question of how 
many supersymmetric and non-supersymmetric fixed points are encountered at 
intermediate scales in the space of couplings before a desirable low-energy 
fixed point is reached is, of course, model dependent. To our knowledge, this 
type of behavior in supersymmetric theories has not been noticed prior to this 
work and whether similar behavior occurs in higher dimensional theories is an 
important question which could prove relevant to the general study of 
supersymmetry (breaking) in phenomenologically interesting models and deserves 
further investigation.   

\begin{center}
ACKNOWLEDGEMENTS
\end{center}
This work was supported by the Swedish Natural Science Research Council.

\bibliographystyle{plain}

\end{document}